\documentclass[aps,prc,unsortedaddress,reprint,nofootinbib]{revtex4-1}

\usepackage{epsfig}
\usepackage{dcolumn}
\usepackage{bm}
\usepackage{amssymb}
\usepackage{amsmath}
\usepackage{hyperref}

\usepackage{graphicx}
\usepackage[dvipsnames]{xcolor}
\usepackage{color}
\definecolor{darkblue}{rgb}{0.0, 0.0, 0.55}
\definecolor{cite}{rgb}{0.0, 0.34, 0.25}

\usepackage{rotating}
\usepackage{bigints}
\usepackage{lineno}

\newcommand{\MeV}{\, {\rm MeV}}
\newcommand{\GeV}{\, {\rm GeV}}
\newcommand{\TeV}{\, {\rm TeV}}
\newcommand{\fm}{\, {\rm fm}}

\begin{document}
\title{Influence of hadronic resonances on the chemical freeze-out in heavy-ion collisions}

\author{P. Alba}
\affiliation{Lucht Probst Associates GmbH, Grosse Gallusstrasse 9, D-60311 Frankfurt am Main, Germany}

\author{V. Mantovani  Sarti}
\affiliation{Physik Department T70, E62, Technische Universit\"at M\"unchen, 
James Franck Strasse 1, 85748 Garching, Germany}

\author{J. Noronha-Hostler}
\affiliation{Department of Physics, 
University of Illinois at Urbana-Champaign, Urbana, IL 61801, USA}

\author{P. Parotto}
\affiliation{Department of Physics, University of Houston, Houston, TX 77204, USA}
\affiliation{University of Wuppertal, Department of Physics, Wuppertal D-42119, Germany}

\author{I. Portillo-Vazquez}

\author{C. Ratti}

\author{J. M. Stafford}
\email[Corresponding author: ]{jmstafford@uh.edu}
\affiliation{Department of Physics, University of Houston, Houston, TX 77204, USA}

\date{\today}

\begin{abstract}
Detailed knowledge of the hadronic spectrum is still an open question, which has
phenomenological consequences on the study of heavy-ion collisions. A previous lattice
QCD study concluded that additional strange resonances are missing
in the currently tabulated lists provided by the Particle Data Group (PDG). That study
identified the list labeled PDG2016+ as the ideal spectrum to be used as an input in
thermal-model-based analyses. In this work, we determine the effect of additional resonances
on the freeze-out parameters of systems created in heavy-ion collisions. These
parameters are obtained from thermal fits of particle yields and net-particle fluctuations.
For a complete picture, we compare several hadron lists including both experimentally
discovered and theoretically predicted states. We find that the inclusion of additional
resonances mildly influences the extracted parameters -- with a general trend of
progressively lowering the temperature -- but is not sufficient to close the gap in
temperature between light and strange hadrons previously observed in the literature.
\end{abstract}

\maketitle

\section{Introduction}

Determining the number of hadronic resonances and their corresponding properties has
been a fundamental question in nuclear physics for decades. Experimentally, new
resonances (and their decay channels) are measured through spectroscopy and an up-to-date compilation of those searches can be found in the Particle Data Group booklet \cite{Tanabashi:2018oca}. Each new particle is assigned a star on a scale of * to **** 
which indicates the experimental confidence of the measurement, where * states 
are the most uncertain and **** states are the most established. Theoretical 
calculations, such as Quark Models \cite{Capstick:1986bm,Ebert:2009ub}, predict many
more resonances beyond those experimentally measured.  Furthermore, there have been
general theoretical arguments, originally made by Hagedorn, that near a limiting
temperature (now understood as the phase transition of the hadron gas into the Quark
Gluon Plasma \cite{Cabibbo:1975ig}) an exponentially increasing mass spectrum
\cite{Hagedorn:1965st} is expected,  i.e. heavier and heavier resonances can form that
decay almost immediately into their daughter particles. Using these so-called ``Hagedorn
states" and comparing thermodynamic quantities to lattice
Quantum Chromodynamics (QCD) calculations, initial evidence was found that the older versions of
the PDG were insufficient to describe the lattice results  \cite{NoronhaHostler:2009cf,Majumder:2010ik,NoronhaHostler:2012ug}. 
With the particle production results published by the ALICE Collaboration 
(${\rm Pb-Pb} \, \, 2.76 \TeV$ \cite{ABELEV:2013zaa,ALICE:2017jyt} and preliminary ${\rm Pb-Pb} \, \, 5.02 \TeV$
\cite{Bellini:2018khg}) a new issue arose, the so-called ``proton anomaly'' wherein thermal model
fits generally over-predict the proton yields while simultaneously under-predicting the
yields of strange baryons. Fluctuation analyses show the same trend 
\cite{Bellwied:2018tkc,Bluhm:2018aei,Bellwied:2019pxh}. A number of possible solutions
were suggested such as a flavor hierarchy of freeze-out temperatures between light and
strange hadrons \cite{Bellwied:2013cta}, further missing Hagedorn states 
\cite{Bazavov:2014xya,Noronha-Hostler:2014usa,Noronha-Hostler:2014aia}, final state
reactions \cite{Rapp:2000gy,Steinheimer:2012rd}, in-medium effects \cite{Aarts:2018glk} or corrections to the ideal HRG model 
\cite{Lo:2017lym,Andronic:2018qqt}.

In order to settle much of these debates on the number of hadron resonances, lattice
calculations have played a vital role in recent years. Initial results from Ref. 
\cite{Bazavov:2014xya} found evidence of missing strange resonances from first principle
calculations. In a more recent work, using partial contributions to the QCD pressure of
particles grouped according to their quantum numbers, the exact nature of the missing
resonances was determined \cite{Alba:2017mqu}. Evidence for the need of additional
states was seen in certain sectors and even with their inclusion, the contribution of
strange mesons still underestimates lattice QCD results. However, in Ref. 
\cite{Alba:2017mqu} it was shown that the Quark Model predicts too many states -- in
particular in the $\left|S\right| = 2$ sector -- and thus overestimates most of the partial
pressures. The list which correctly reproduces the largest number of observables is the
PDG2016+ list (see definition below). For completeness, in this manuscript we will also
consider Quark Model states, in order to investigate the effect of including the largest
possible number of resonances in the analyses.

It is worth mentioning that missing hadronic resonances have been found to affect a
number of areas in heavy-ion collisions such as the transport coefficients $\eta/s$ and 
$\zeta/s$  \cite{NoronhaHostler:2008ju,NoronhaHostler:2012ug,Rais:2019chb}, and may
somewhat affect collective flow \cite{Noronha-Hostler:2013rcw,Alba:2017hhe} and mildly
improve the $\chi^2$ of thermal fits \cite{NoronhaHostler:2009tz}. In Ref. 
\cite{Alba:2017hhe} we incorporated these missing states from the PDG into a
hydrodynamic model and found that they are able to improve the fits to particle spectra
and $\langle p_T\rangle$. Similar results were found in Ref. \cite{Devetak:2019lsk}.

In this follow-up paper, we compare different hadronic lists with an increasing number of
states, performing fits of particle yield and fluctuation data from RHIC and the LHC. One of the
most pressing questions is whether these new states can close the gap between light and
strange freeze-out temperatures, as suggested in Ref. \cite{Bazavov:2014xya}. Initial
explorations in this direction can be found in Refs. \cite{Chatterjee:2013yga,Chatterjee:2017yhp}, 
where smaller particle resonance lists were used. We study a scenario with a freeze-out
temperature common to all species, as well as one where different freeze-out
temperatures result for strange and light species. In the case of fits to yields and their
ratios we compare the fit quality from both scenarios. 

This manuscript is organized as follows. In Section~\ref{sec:hadron_lists} 
we describe the different hadronic lists we utilize in this work. Moreover, we provide 
the details on how the decay properties of the resonant states were estimated, when 
not known experimentally. In Section~\ref{sec:HRG_fits} we briefly describe the HRG model and the data sets employed in the fits we perform. In Sections~\ref{sec:fits_1FO} and~\ref{sec:fits_2FO} we show our results for fits to yields and ratios in the single and double freeze-out scenarios, respectively. In Section~\ref{sec:fits_fluc} we study the chemical freeze-out from an analysis of net-particle fluctuations from the Beam Energy Scan (BES) at RHIC. We first analyze fluctuations of net-proton and net-charge, then those of net-kaon. Finally, in Section~\ref{sec:concl} we present our conclusions.

The particle lists PDG2005, PDG2016, PDG2016+ and QM utilized in this work, together with the
corresponding decay channels can be downloaded from the link provided in 
Ref.~\cite{pdglists:2020}.

\section{Different hadron lists}\label{sec:hadron_lists}

We consider the following lists of hadronic resonances. PDG2005: PDG from
2005 \cite{Eidelman:2004wy}; PDG2016: PDG **-**** states from 2016 
\cite{Patrignani:2016xqp}; PDG2016+: PDG *-**** states from 2016 
\cite{Patrignani:2016xqp}; and QM: PDG2016+ combined with additional Quark Model states
from Refs. \cite{Capstick:1986bm,Ebert:2009ub}. The comparison between PDG2005 and
PDG2016 is performed to demonstrate the difference between earlier thermal fit models
and modern day ones.  The PDG2016+ is tested in thermal fits in this paper for the first
time.  One primary reason why this has not yet been done in the past is that many of the *
and ** states do not have adequate decay channel and branching ratio information in
order to describe all the particle interactions based only on experimental data. In this
paper, we use phenomenology to fill in these gaps. For completeness, we consider a mixture of the
PDG2016+ with the addition of Quark Model states that in Ref. \cite{Alba:2017mqu}
appeared to fit some lattice QCD data well. Note that the list we label as QM does not
exclusively contain states predicted by the Quark Model. We replaced the Quark Model
states that are already known experimentally with the corresponding ones from the
PDG2016+ list. In order to discern whether a state predicted by the Quark Model was
already listed in the PDG, we considered states sharing the same quantum numbers
(baryon number, strangeness, electric charge, isospin and spin degeneracy), with
comparable masses (mass difference below $\sim 150 \, \MeV$, or within $15\%$ of the
mass of the PDG state, whichever is smaller). The issue with the decays is even more
pronounced for the QM, where no decay channels are given. In this case we relied entirely
on phenomenological approaches that are described in detail below.

\subsection{Decay lists}

In the PDG database \cite{Patrignani:2016xqp}, only the relatively few established states
are provided with an exhaustive set of decays, whereas most states are only given
branching ratios which do not sum up to $100\%$. In this case, additional decays are listed as ``dominant", ``seen" or ``not seen". 

When considering the effect of resonance decays, we include all the strong
decays listed by the PDG with a branching ratio of at least $\simeq 1\%$, but
we do not include weak decays. When the listed branching ratios do not sum up to
$100\%$, we assign the remaining percentage to an electromagnetic decay of the 
kind $N_2 \rightarrow N_1 + \gamma$ -- where $N_2$ and $N_1$ are hadrons 
with the same quantum numbers, $N_1$ being the next state in 
descending mass order with compatible parity for such a decay. We decay resonances that lack quantitative information predominantly radiatively (as explained above), in order to populate the next state with known branching ratios.

The situation is different for the hadronic resonant states predicted by relativistic quark
models, since these are not supplied with any information on their decay properties. 
In order to take into account the effect of these additional states on observables, 
we need to estimate their branching ratios.

Along the lines of the work done for the PDG states, we infer the branching ratios of the
additional QM states using the following procedure. Every state is assigned a predominant radiative decay, analogously to PDG states with incomplete information. The remaining percentage is reserved for hadronic decays, for which we use linear extrapolation from known branching ratios of PDG particles with the same quantum numbers $B,~S,~I$ and $Q$ (baryon number, strangeness, isospin and electric charge). 

The linear extrapolation is performed as follows. States from the PDG2016+ list are
divided into ``families'' with the same quantum numbers, and all strong decay
modes present in the family are grouped together. For each channel appearing in 
the family, a linear interpolation of its branching ratio as a function of the particle mass 
is performed. Then, QM states of the same family are assigned branching ratios 
by sampling the linear mass dependence constructed as explained above. At this point, 
all negative BR values are discarded, as well as all decays that violate 
mass conservation. Finally, the sum of branching ratios for strong decays is normalized and put together with the electromagnetic ones. 

We emphasize here that for the QM states we have the least amount of information and,
therefore, the largest amount of uncertainties.

\section{Thermal Fits and the Hadron Resonance Gas model}\label{sec:HRG_fits}

The study of chemical freeze-out we present here makes use of the Hadron Resonance
Gas (HRG) model, which describes the hadronic phase below the transition temperature
$T_c$ as a system of non-interacting particles~\cite{Hagedorn:1965st,Dashen:1969ep,Venugopalan:1992hy}. 
The HRG model has a wide-spread applicability in heavy-ion studies in reproducing
thermal particle abundances and lately also in providing results on fluctuations of
conserved charges $(B,~Q,~S)$ \cite{Alba:2014eba,Alba:2015iva,Bellwied:2018tkc,Bluhm:2018aei,Poberezhnyuk:2019pxs}. Recently, these observables 
have been compared to the measured moments of net-particle distributions, and provided freeze-out
temperatures which are compatible with the ones obtained by comparing the experimental data to lattice QCD calculations 
\cite{Alba:2014eba,Karsch:2010ck,Garg:2013ata,FU2013144,Borsanyi:2013hza,Borsanyi:2014ewa}.

Historically, the HRG model has been widely employed to compare data on particle
production for energies ranging from the AGS to the LHC~\cite{Becattini:2005xt,Becattini:2012xb,Cleymans:2005xv,Torrieri:2006yb,Andronic:2011yq,BraunMunzinger:2003zd,Andronic:2005yp,Bhattacharyya:2019wag,Bhattacharyya:2019cer}.
Produced particle yields $\langle N_i\rangle$ are obtained by adding the contribution from resonances
to the primordial thermal yield, given by $Vn_i$: 
\begin{equation}
\label{Nh}
\langle N_i \rangle ~=~V\,n_i~+~V\,\sum_R\langle n_i\rangle_R\,n_R~.
\end{equation}
In the above, $\langle n_i\rangle_R$ is the average number of particles of type $i$ resulting 
from a decay of resonance $R$,  $n_i$ and $n_R$ are thermal densities calculated 
through the statistical model, and $V$ is the system volume.
Conditions on the net-strangeness and net-charge density are imposed, to match the heavy-ion collision situation:
\begin{align}
& \langle n_S(T;\mu_B,\mu_Q,\mu_S)\rangle=0\nonumber \, \, ,\\
& \langle n_Q(T;\mu_B,\mu_Q,\mu_S)\rangle=\dfrac{Z}{A} \langle n_B\rangle \, \, .
\end{align}
These allow one to constrain the three chemical potentials. In this way, yields and ratios
calculated within the HRG model only depend on the thermal parameters $(T,\mu_B)$
(and $V$ in the case of yields).

Thermal properties at the chemical freeze-out have been studied using yields and 
ratios from STAR data in Au-Au collisions at $\sqrt{s_{\mathrm{NN}}}=200,~130,~39,~27,~19.6,~11.5,~7.7$ GeV~\cite{Abelev:2008ab,Adams:2006ke,Adamczyk:2017iwn,Cleymans:2004pp} and from ALICE data in Pb-Pb collisions at $\sqrt{s_{\mathrm{NN}}}=2.76 \TeV$
and $5.02 \TeV$ \cite{Abelev:2013vea,ABELEV:2013zaa,Abelev:2013xaa,Abelev:2014uua,Andronic:2017pug,Bellini:2018khg}.

In this manuscript, we focus on the STAR data at $\sqrt{s_{\mathrm{NN}}}=200$ GeV 
and $0-5\%$ centrality, and the ALICE data at $\sqrt{s_{\mathrm{NN}}}=5.02$ TeV 
and $0-10\%$ centrality. We perform thermal fits of the particle yields and ratios, 
using published data on ratios, if available, for STAR~\cite{Abelev:2008ab} and 
for ALICE. In order to build 
the remaining ratios, published data on yields from both collaborations have been 
used with a proper propagation of the errors in the final result.
We evaluate the yields and ratios for each hadronic list and extract the thermal
parameters $(T,~\mu_B,~V)$ by using the thermal fit package FIST \cite{Vovchenko:2019pjl}. The package allows users to choose their own particle lists, as well as data sets, 
in the fit.

\section{Single Freeze-out Scenario}\label{sec:fits_1FO}

Initially, we consider a common freeze-out temperature for strange and light hadrons. We fit the measured yields for the following particles:
$\pi^+,~\pi^-,~K^+,~K^-,~p,~\bar{p},~\Lambda,~\bar{\Lambda},~\phi,~\Xi,~\bar{\Xi},~\Omega,~\bar{\Omega}$ at the LHC, while at RHIC the separate $\Omega$ and $\bar{\Omega}$ yields are replaced by the sum $\Omega+\bar{\Omega}$. When we take the ratios, we divide the light particle yields by the yield of either $\pi^+$ or $\pi^-$ and the strange particle yields by the yield of either $K^+$ or $K^-$. The
results of the thermal fits for both yields and ratios while varying the particle resonance 
list are shown in Table~\ref{tab:1T_ALICE} for LHC data at $\sqrt{s_{NN}} = 5.02 \TeV$ 
and  in Table~\ref{tab:1T_STAR} for STAR data at $\sqrt{s_{NN}} = 200 \GeV$. At the LHC 
we hold $\mu_B = 1 \MeV$ fixed to avoid the possibility of negative chemical potentials 
and remain consistent with previous analyses \cite{Andronic:2012dm}.

\begin{table*}
\begin{tabular}{| c | c | c | c | c | c | c | c |}
\hline
 & \multicolumn{2}{c|}{T [MeV]} & \multicolumn{2}{c|}{$\mu_B$ [MeV]} & Volume [$\fm^3$]  & \multicolumn{2}{c|}{$\chi^2$/DOF} \\
 \hline
 &Yields &Ratios  &Yields &Ratios &  Yields &Yields &Ratios \\
\hline
PDG2005   &  158.3 $\pm$ 1.8  & $155.7\pm1.6$ & 1 & 1 & 4904 $\pm$ 456 & 74.0/11 & 80.7/12 \\
\hline
PDG2016   &152.1 $\pm$ 1.7 & $149.2\pm1.4$ & 1 & 1 & 5811 $\pm$ 545 & 102.4/11  & 111.5/12 \\
\hline
PDG2016+  &150.4 $\pm$ 1.5 & $148.1\pm1.3$ & 1 & 1 & 6250 $\pm$ 561 & 79.0/11  & 89.0/12 \\
\hline
QM      &147.9 $\pm$ 1.4 & $146.0\pm1.2$ & 1 & 1 & 6867 $\pm$ 581 & 64.7/11  & 75.0/12 \\
\hline
\end{tabular}
\caption{Temperatures and baryon chemical potential (fixed to 1 MeV) in the single chemical freeze-out 
scenario. The results are obtained from thermal fits to total yields and ratios from ALICE 
data (\cite{Bellini:2018khg}) at $0-10\%$ centrality in PbPb collisions, at 
$\sqrt{s_{NN}} = 5.02 \TeV$, using different particle lists. For the case of total yields, the 
volume is shown as well. The last two columns show the chi square per number of 
degrees of freedom in the fits.}
\label{tab:1T_ALICE}
\end{table*}

\begin{table*}
\begin{tabular}{| c | c | c | c | c | c | c | c |}
\hline
 & \multicolumn{2}{c|}{T [MeV]} & \multicolumn{2}{c|}{$\mu_B$ [MeV]} & Volume [$\fm^3$] & \multicolumn{2}{c|}{$\chi^2$/DOF} \\
 \hline
 &Yields &Ratios  &Yields &Ratios & Yields &Yields &Ratios \\
\hline
PDG2005   &160.2 $\pm$ 2.1& $164.0\pm2.7$ & 23.2 $\pm$ 8.1 & $24.4\pm9.3$ & 2251 $\pm$ 260 & 42.9/9 & 15.6/10\\
\hline
PDG2016   &159.8 $\pm$ 2.3 & $161.6\pm2.8$ & 22.5 $\pm$ 7.2 & $24.1\pm8.6$ & 1984 $\pm$ 270  & 32.0/9 & 16.3/10\\
\hline
PDG2016+  &156.4 $\pm$ 2.2 & $158.5\pm2.5$ & 21.1 $\pm$ 6.6 & $23.1\pm8.3$ & 2248 $\pm$ 304  & 14.5/9 & 8.7/10\\
\hline
QM      &151.5 $\pm$ 1.8 & $154.0\pm2.1$ & 20.3 $\pm$ 6.6 & $22.4\pm8.2$ & 2768 $\pm$ 333  & 18.4/9 & 8.8/10\\
\hline
\end{tabular}
\caption{Temperatures and baryon chemical potentials from thermal fits with a single 
chemical freeze-out scenario compared to either ratios or total yields from STAR 
data~\cite{Abelev:2008ab,Adams:2006ke} for $0-5\%$ centrality in AuAu collisions at 
$\sqrt{s_{NN}} = 200 \GeV$, using different particle lists. For the case of total yields, the 
volume is shown as well. The last two columns show the chi square per number of 
degrees of freedom in the fits.}
\label{tab:1T_STAR}
\end{table*} 
From Tables \ref{tab:1T_ALICE} and \ref{tab:1T_STAR} a few trends begin to emerge:
\begin{itemize}
\item for both yields and ratios, more hadronic states generally decrease the chemical freeze-out temperature;
\item the chemical freeze-out temperatures from yields and ratios approximately agree; 
\item generally, due to the decrease of the chemical freeze-out temperature with the increase of the number of hadronic states, the volume increases accordingly; 
\item $\mu_B$ is relatively stable when changing the particle resonance list;
\item RHIC appears to have somewhat higher chemical freeze-out temperatures compared to the LHC;
\item the inclusion of resonant states beyond the PDG2016+ list improves the fit quality at the LHC for both yields and ratios, while at RHIC the PDG2016+ fits best for the yields and the QM and PDG2016+ are equivalent for the ratios.
\end{itemize}

\begin{figure}
\begin{center}
\includegraphics[width=0.98\linewidth]{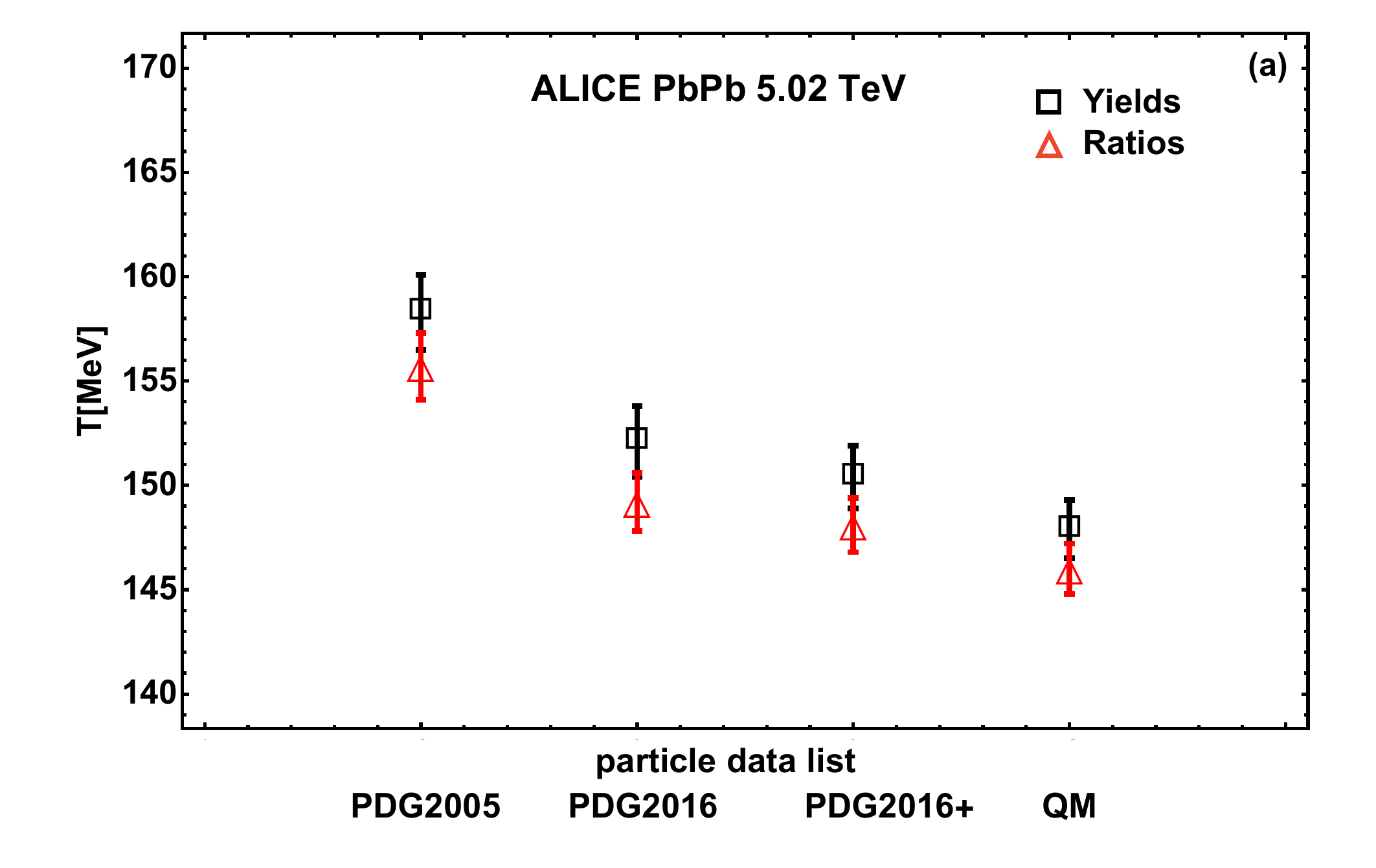}
\includegraphics[width=0.98\linewidth]{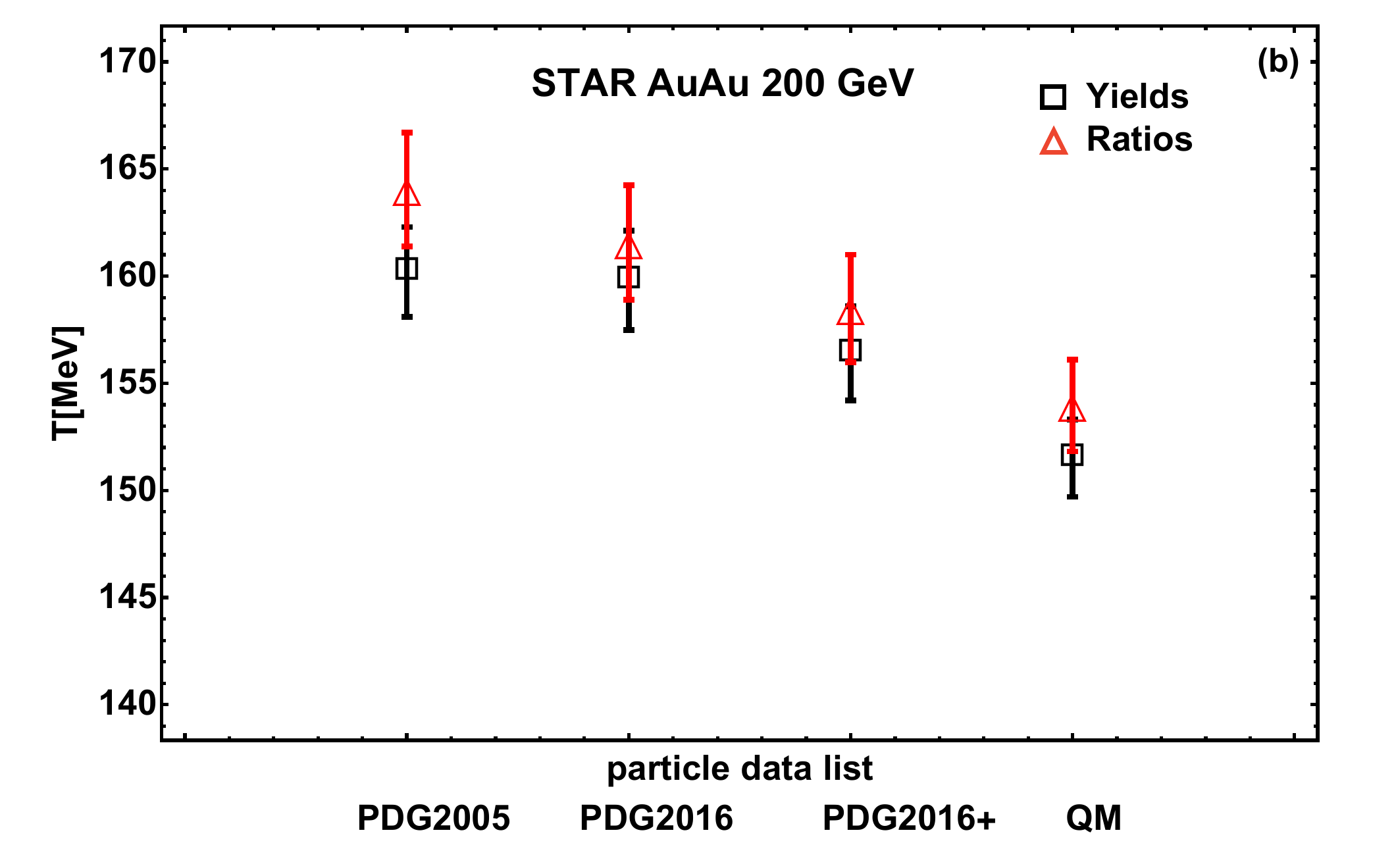}
\caption{
Dependence of the chemical freeze-out temperature on the particle data lists at the LHC 
(upper panel) and RHIC (lower panel). We compare the results extracted 
from yields (black squares) and ratios (red triangles).
} 
\label{Tchmone}
\end{center}
\end{figure}

\begin{figure}
\begin{center}
\includegraphics[width=0.98\linewidth]{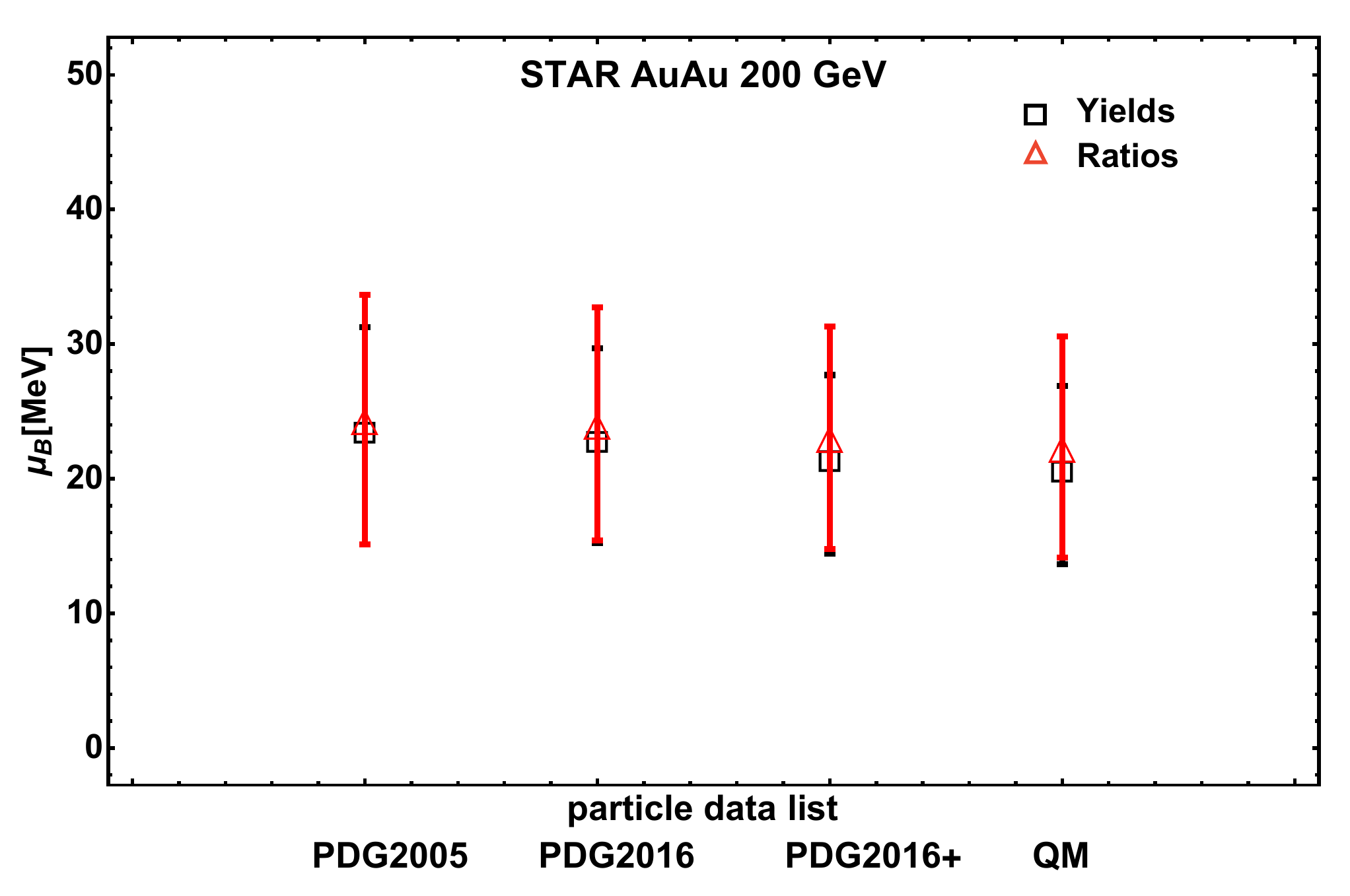}
\caption{
Dependence of $\mu_B$ on the particle data lists at RHIC. We compare the results 
extracted from yields (black squares) and the ratios (red triangles).
} 
\label{MUBchmone}
\end{center}
\end{figure}

The second to last point is puzzling because one expects the freeze-out curves to 
remain roughly constant at low baryon chemical potential, according to lattice QCD 
studies of the phase transition at finite $\mu_B$~\cite{Bellwied:2015rza,Bazavov:2017tot}. 

The values of temperature and chemical potential listed in 
Tables~\ref{tab:1T_ALICE} and \ref{tab:1T_STAR} are shown in 
Fig.~\ref{Tchmone} and Fig.~\ref{MUBchmone}, respectively.

In Fig.~\ref{Tchmone}, the decreasing trend of the chemical freeze-out temperature with 
the increasing number of hadronic states in the list is clearly visible. As mentioned previously, 
slightly higher temperatures are found at RHIC. 
On the other hand, in Fig.~\ref{MUBchmone} we can see that the RHIC freeze-out chemical 
potential barely changes with different resonance lists, and that the results
from yields and ratios agree well.

\section{Two Freeze-out Scenario}\label{sec:fits_2FO}

\begin{table*}
\begin{tabular}{| c | c | c | c | c | c | c | c | c |}
\hline
 & & \multicolumn{2}{c|}{T [MeV]} & \multicolumn{2}{c|}{$\mu_B$ [MeV]} & Volume [$\fm^3$]  & \multicolumn{2}{c|}{$\chi^2$} \\
 \hline
 & &Yields &Ratios  &Yields &Ratios &  Yields &Yields &Ratios \\
\hline
PDG2005 & Light  &148.8 $\pm$ 2.2 & 144.9 $\pm$ 2.0 & 1 & 1 & 7598 $\pm$ 837 & 17.1/4 & 0.058/5 \\
\hline
 & Strange  &167.5 $\pm$ 2.2 & 167.5 $\pm$ 2.0 & 1 & 1 & 3212 $\pm$ 335 & 23.8/7 & 20.1/8 \\
\hline
PDG2016 & Light  &143.2 $\pm$ 1.8 & 140.3 $\pm$ 1.6 & 1 & 1 & 9096 $\pm$ 897 & 15.8/4  & 0.062/5 \\
\hline
 & Strange  &169.8 $\pm$ 2.6 & 169.8 $\pm$ 2.3 & 1 & 1 & 2472 $\pm$ 319 & 7.2/7  & 6.1/8 \\
\hline
PDG2016+ & Light &142.5 $\pm$ 1.7 & 139.8 $\pm$ 1.6 & 1 & 1 & 9374 $\pm$ 902 & 14.7/4  & 0.063/5 \\
\hline
& Strange &165.0 $\pm$ 2.4 & 164.7 $\pm$ 2.1 & 1 & 1 & 2978 $\pm$ 377 & 0.77/7  & 0.70/8 \\
\hline
QM & Light &140.9 $\pm$ 1.6 & 138.5 $\pm$ 1.5 & 1 & 1 & 9961 $\pm$ 927 & 14.0/4  & 0.063/5 \\
\hline
 & Strange &158.5 $\pm$ 1.9 & 158.5 $\pm$ 1.7 & 1 & 1 & 3835 $\pm$ 433 & 0.12/7 & 0.10/8 \\
\hline
\end{tabular}
\caption{Temperatures and baryon chemical potential (fixed to 1 MeV) in the double chemical freeze-out 
scenario obtained from fits to ratios or total yields from ALICE 
data~\cite{Bellini:2018khg} for 
$0-10\%$ centrality in PbPb collisions at $\sqrt{s_{NN}}=5.02$ TeV, using different PDG 
lists. For the case of total yields, the volume is shown as well. The last two columns show the chi square per number of 
degrees of freedom in the fits.}
\label{tab:2T_ALICE}
\end{table*}

\begin{table*}
\begin{tabular}{| c | c | c | c | c | c | c | c | c |}
\hline
 & & \multicolumn{2}{c|}{T [MeV]} & \multicolumn{2}{c|}{$\mu_B$ [MeV]} & Volume [$\fm^3$]  & \multicolumn{2}{c|}{$\chi^2$} \\
 \hline
 & &Yields &Ratios  &Yields &Ratios &  Yields &Yields &Ratios \\
\hline
PDG2005 & Light  & 162.4 $\pm$  7.0 & $159.5 \pm 6.2$ & $23.9\pm 14.6$ & $24.1\pm 10.7$  & 1943 $\pm$ 589 & 3.1/3 & 0.12/4 \\
\hline
 & Strange  & 160.0 $\pm$  2.6 & $165.0\pm 2.9$ & $23.1\pm 9.8$ & $24.3\pm 18.3$ & 2284 $\pm$ 362 & 42.8/5 & 14.8/6 \\
\hline
PDG2016 & Light  & 153.6 $\pm$  5.1 & $151.8\pm 4.7$ & $22.6\pm 14.1$ & $23.2\pm 10.3$ & 2528 $\pm$ 650 & 3.1/3  & 0.16/4 \\
\hline
 & Strange  & 161.0 $\pm$  3.0 & $166.1\pm 3.2$ & $21.3\pm 8.9$ & $22.4\pm 17.1$ & 1880 $\pm$ 352 & 28.4/5  & 10.1/6 \\
\hline
PDG2016+ & Light & 152.5 $\pm$  4.9 & $150.9\pm 4.5$ & $22.4\pm 13.9$ & $23.0\pm 10.2$ & 2631 $\pm$ 654 & 2.9/3  & 0.18/4 \\
\hline
& Strange & 158.4 $\pm$  3.1 & $161.7\pm 2.9$ & $20.3 \pm 7.8$ & $21.0\pm 14.7$ & 1981 $\pm$ 403 & 12.7/5 & 4.5/6 \\
\hline
QM & Light & 150.1 $\pm$  4.5 & $148.8\pm 4.1$ &  $22.0\pm 13.8$ & $22.7\pm 10.1$ & 2860 $\pm$ 679 & 2.9/3  & 0.23/4 \\
\hline
 & Strange & 152.4 $\pm$  2.5 & $155.8\pm 2.4$ & $19.5\pm 7.8$ & $20.3\pm 14.8$ & 2600 $\pm$ 474 & 17.8/5  & 6.5/6 \\
\hline
\end{tabular}
\caption{Temperatures and baryon chemical potentials in the double chemical freeze-out 
scenario obtained from fits to ratios or total yields from STAR data~\cite{Abelev:2008ab,Adams:2006ke} 
for $0-5\%$ centrality in AuAu collisions at $\sqrt{s_{NN}}=200$ GeV, using different PDG 
lists. For the case of total yields, the volume is shown as well. The last two columns show the chi square per number of 
degrees of freedom in the fits.}
\label{tab:2T_STAR}
\end{table*} 

We now explore the possibility of a flavor hierarchy between light and strange particles, 
using the same particle resonance lists as in the previous section. There, we saw
that the Quark Model list produces the best fit to the experimental data at ALICE, while at RHIC the PDG2016+ provided the best fit for the yields and for the ratios the QM and PDG2016+ list were nearly equivalent. Nevertheless, we stress here again that the QM results are shown here just for illustration, as in Ref. \cite{Alba:2017mqu} it was found that the QM contains too many states and fails to reproduce several thermodynamic quantities from lattice QCD.
For this analysis, we allow for two separate chemical freeze-out 
temperatures as well as two separate freeze-out baryon chemical potentials for light and strange particles.

For the light hadron fit we use $\pi^+$, $\pi^-$, $p$, $\overline{p}$, $K^+$ and $K^-$, while for the 
strange hadrons fit we use all species except pions and (anti)protons. We construct the 
ratios of strange hadrons dividing by the $K^\pm$ yields, to avoid mixing of light and strange degrees of freedom. We do this also because, if light particles freeze-out at a lower temperature than the one for strange particles, the respective volumes would be different and thus would not cancel out when taking ratios. In 
Tables~\ref{tab:2T_ALICE} and~\ref{tab:2T_STAR} the extracted chemical freeze-out 
temperatures, chemical potentials and volumes (for yields only, as before) are shown for 
both the light and strange hadrons at the LHC and RHIC, respectively. As in the case of a 
single freeze-out scenario, there is very good agreement between fits to yields and ratios. 
Moreover, the freeze-out temperature for strange hadrons is systematically higher than 
that of light hadrons.  

An additional effect of the two temperatures is that the light temperature is significantly 
lower than in the single freeze-out temperature scenario. This means that the fit in that 
case is driven by the strange states, as it was discussed in \cite{Alba:2015iva}. We also 
notice that the inclusion of more resonances affects the strange freeze-out temperature 
more than the light one. This is not unexpected, as most of the additional states carry 
strangeness.

\begin{figure*}
\includegraphics[width=\textwidth]{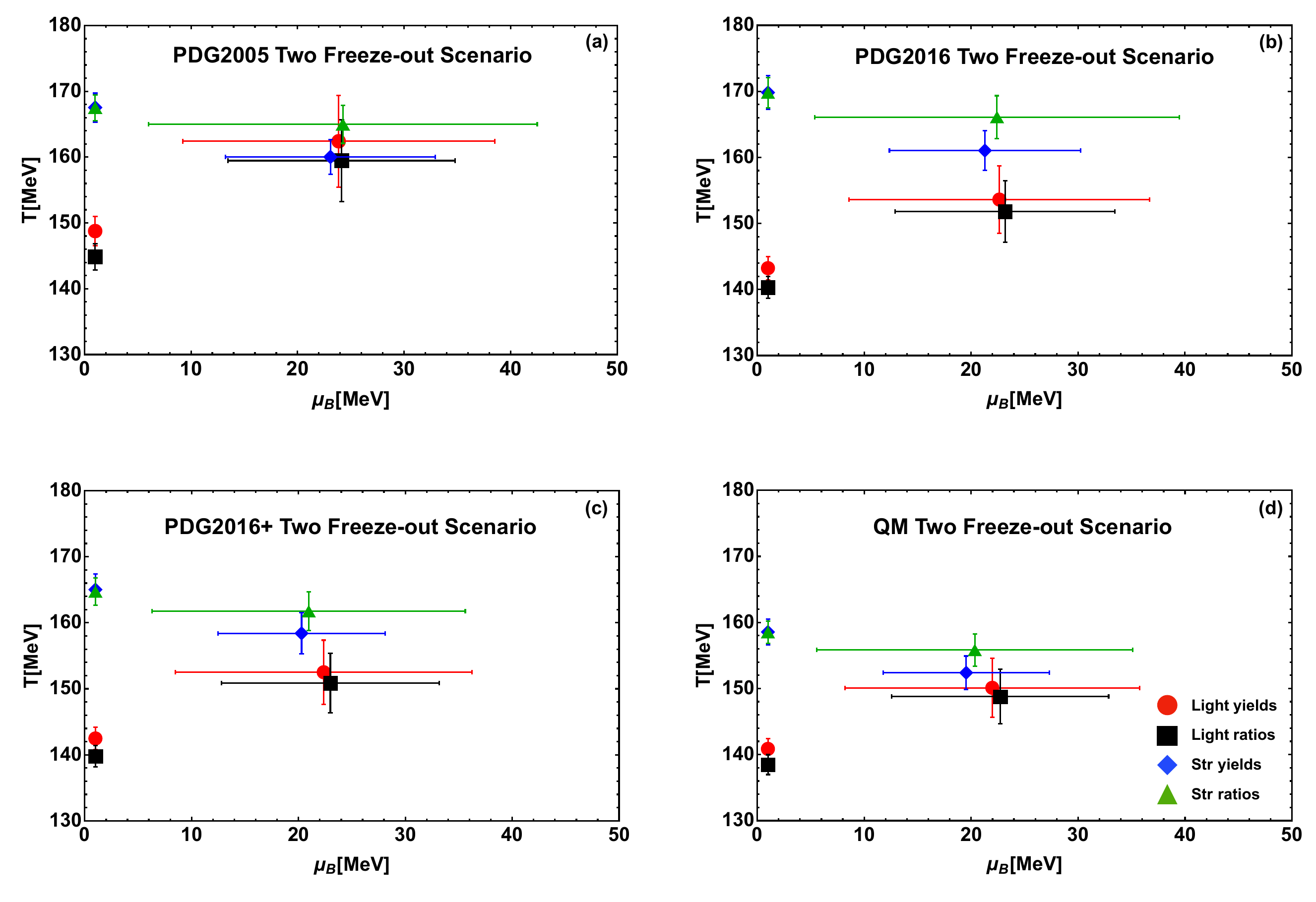}

\caption{
(Color online) Extracted light and strange freeze-out temperatures and chemical 
potentials from yields and ratios at the LHC and RHIC, in the case of a double freeze-out 
scenario. Each panel corresponds to a different particle list.
} 
\label{fit-yields_all}
\end{figure*}

A summary of our results from Tables~\ref{tab:2T_ALICE} and \ref{tab:2T_STAR} is shown in Fig.~\ref{fit-yields_all}, for
yields and ratios from both ALICE and STAR, and with the four different hadron lists we consider. 
In this pictorial representation, the separation in temperature between strange and light 
freeze-out appears very clearly, especially from fits to LHC data. In this case, all lists lead to a 
very pronounced separation. In the case of STAR data, the separation is especially evident 
when utilizing the PDG2016 and PDG2016+ lists, although still present even for fits with the 
QM list. There seems to be a general trend in these last three hadron lists, where the 
presence of more states leads to a smaller separation of the strange and light freeze-out 
temperatures. The PDG2005 list is excluded from this trend: it contains a significantly
smaller number of resonances, which is not realistic in modern calculations. On the other 
hand, we also know that the strange states contained in the QM list are too numerous, as 
noted earlier and shown in Ref.~\cite{Alba:2017mqu}.

In Appendix~\ref{app:graphs} (Figs.~\ref{fit-yields_lhc1}-\ref{fit-yields_star2}) a comparison 
of the yields and ratios to experimental data is shown, both in the single (red lines) and double (blue dotted lines) freeze-out scenarios.

\section{Freeze-out from fluctuations}\label{sec:fits_fluc}

In this Section we complete our analysis by studying net-particle fluctuations from the BES 
at RHIC. We employ the state-of-the-art list PDG2016+ to perform the analysis of net-proton 
and net-charge fluctuations from  Ref.~\cite{Alba:2014eba}, as well as the analysis of 
net-kaon fluctuations from Ref.~\cite{Bellwied:2018tkc}. Both analyses were originally 
carried out with an older PDG list, which we here indicate as PDG2012.

\begin{figure}
\includegraphics[width=\linewidth]{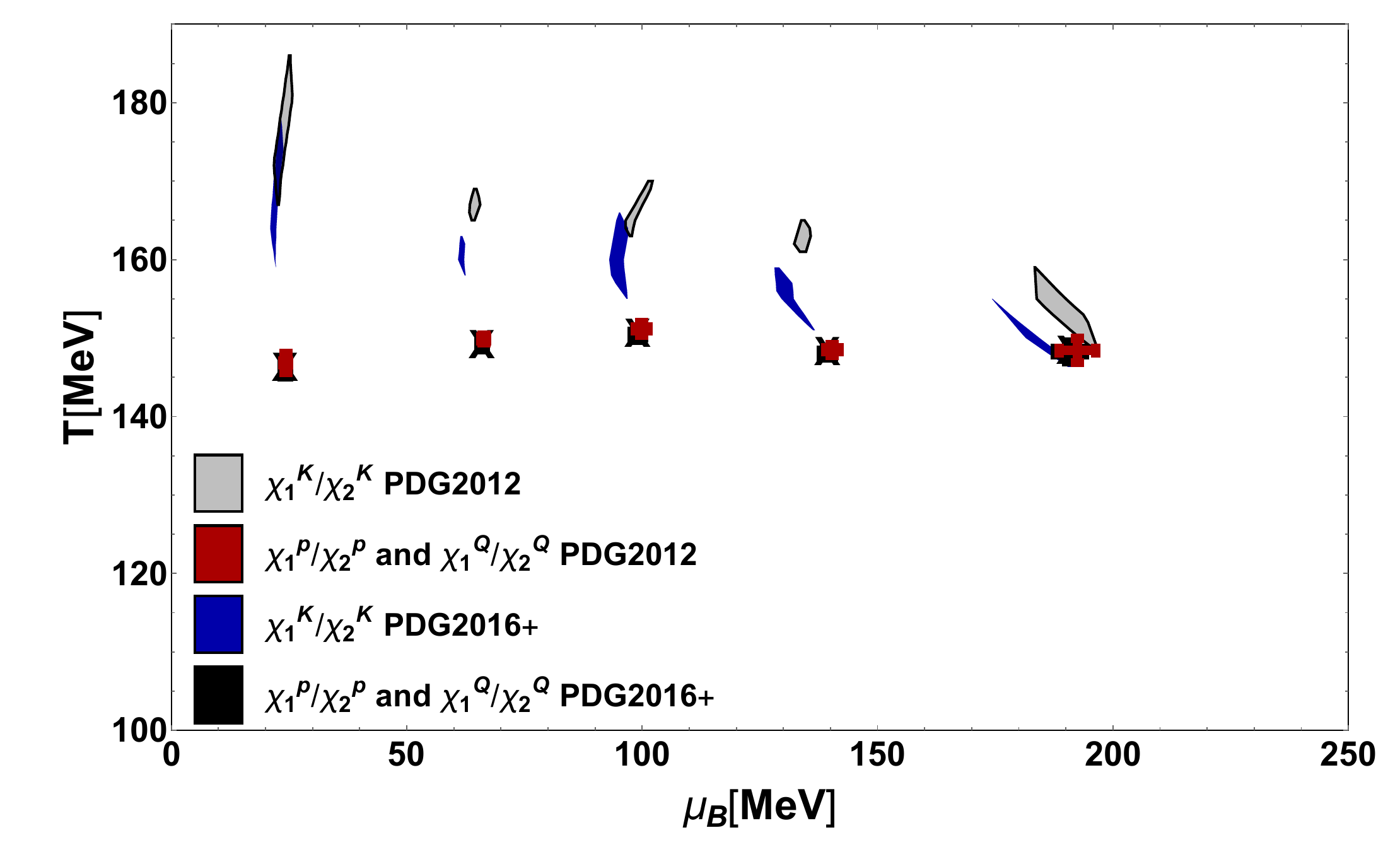}\\
\caption{
(Color online) Comparison of freeze-out parameters obtained via a combined fit of net-proton and net-electric charge with those obtained from the net-kaon fluctuations 
following the analysis from Ref. \cite{Bellwied:2018tkc} with different particle lists.
} 
\label{fig:fluctuations}
\end{figure}

In order to determine both the temperature and chemical potential at the chemical 
freeze-out, in general two experimental quantities are required.
However, due to 
the large experimental errors on higher order fluctuations \ (i.e., the ratio 
$\chi_3/\chi_2$) as noted in \cite{Adam:2020kzk,Koch:2008ia}, in Ref.~\cite{Bellwied:2018tkc} we obtained the freeze-out 
parameters as follows. Starting from the net-proton and net-charge freeze-out 
parameters obtained in Ref.~\cite{Alba:2014eba}, we followed the isentropic trajectories 
determined in Ref.~\cite{Gunther:2016vcp} via a Taylor-expanded equation 
of state from lattice QCD. These were constructed by first determining the entropy-per-baryon ratio at the freeze-out point for each collision energy from Ref.~\cite{Alba:2014eba}, 
and then following the path that conserves $S/N_B$\footnote{Isentropic lines are the 
trajectories followed by the system in the case where no dissipation is present. This is a 
good approximation due to the extremely low viscosity of the medium created in heavy-ion 
collisions, and more so if they are utilized in a small portion of the evolution, in the proximity 
of the transition, as we do here.}. The ratio $\chi^K_1/\chi^K_2$ was then calculated along 
these trajectories , and compared to the experimental results. For each collision energy, 
the overlap region will then correspond to the freeze-out points for net-kaons, 
which are shown in Fig.~\ref{fig:fluctuations} in gray, while the red points correspond 
to the net-proton and net-charge freeze-out points from Ref.~\cite{Alba:2014eba}.

In order to repeat the analysis with the PDG2016+ list, we first calculated
the freeze-out parameters for light hadrons via a combined fit of $\chi_1^p/\chi_2^p$ 
and $\chi_1^Q/\chi_2^Q$ with the new list. We then used these 
light freeze-out parameters to determine 
the new isentropic trajectories in the QCD phase diagram. As was done in 
Ref.~\cite{Gunther:2016vcp}, we utilized a Taylor-expanded equation of state from 
lattice QCD. With the new isentropic trajectories, we could then calculate the 
net-kaon fluctuations along them and determine the freeze-out parameters. 

We note that, both for the net-proton and net-charge, as well as for the net-kaon 
analysis, we included the effect of resonance decays, and imposed the same kinematic cuts 
employed in the experiments \cite{Adamczyk:2013dal,Adamczyk:2014fia,Adamczyk:2017wsl}. Moreover, for the light hadron analysis we included the 
effect of isospin randomization \cite{Kitazawa:2011wh,Kitazawa:2012at}, which has a large 
impact on net-proton fluctuations.

The results for the light and kaon freeze-out with the PDG2016+ list are shown in 
Fig.~\ref{fig:fluctuations} with black points and blue shapes, respectively. We can see 
that, as in the analysis with the PDG2012 list from Refs.~\cite{Alba:2014eba,Bellwied:2018tkc}, there is a separation of the net-kaon from the light particle freeze-out temperature at the higher 
collision energies (lower $\mu_B$) that decreases as the collision energy decreases. 
With the inclusion of more resonances in the HRG model, the 
freeze-out temperature for the net-kaons is decreased, while the light hadron 
freeze-out is mostly unchanged. However, the gap between the two freeze-out 
temperatures is not closed, and the resulting temperatures are compatible with the 
ones obtained in the previous Section with the analysis of fits in the double 
freeze-out scenario.

\section{Conclusions}\label{sec:concl}

In this paper we explored the effects that including additional, not-well-established hadronic 
resonances in the PDG lists, as well as a combination of the latter with predicted (but not yet 
measured) Quark Model states, has on the extracted freeze-out parameters in heavy-ion 
collisions. In Ref.~\cite{Bazavov:2014xya} it was suggested that the addition of missing 
strange states could possibly explain the tension between proton and multi-strange hadrons 
production at the LHC. An alternative picture was proposed in Ref.~\cite{Bellwied:2013cta} 
wherein strange hadrons freeze-out at a higher temperature than the light ones, which is 
due to the higher hadronization temperature of strange particles compared to light 
particles. We explore this alternative picture here. 

For the first time in this paper, we systematically explore the effect of enlarged hadronic resonance spectra 
-- along with their decay properties -- on thermal fits, and find that this tension is not 
resolved. Especially at the LHC, the separation between strange and light freeze-out is 
extremely pronounced. At RHIC the separation is smaller, and is almost resolved when 
states from the Quark Model are included, although it is important to remember that 
certainly too many strange states are predicted by these calculations, as was shown in 
Ref.~\cite{Alba:2017mqu}. From thermal fits based on the most realistic particle list 
PDG2016+, we estimate the light chemical freeze-out temperature to be 
$T_L \sim 141-144$ MeV and the strange chemical freeze-out temperature to be $T_S \sim 163-167$ MeV at the LHC, and $T_L\sim 148-158$ MeV 
and $T_S \sim 155-161$ MeV at RHIC. Generally, the light chemical freeze-out temperature 
appears to be consistent with the net-p and net-Q results (within error bars), while the 
strange temperature is consistent with the one obtained from net-K fluctuations.

Finally, we note that in other works \cite{Steinheimer:2012rd,Noronha-Hostler:2014aia,Noronha-Hostler:2014usa} 
it was suggested the a dynamical scenario could explain the proton to pion puzzle at the 
LHC.  The inclusion of the additional states in such a scenario is left for a future work. 
Generally, we find hints of two separate chemical freeze-out temperatures, but smaller 
error bars in the experimental data would be needed before a definitive answer can be 
given. 

\section*{Acknowledgements}
The authors gratefully acknowledge many fruitful discussions with Rene Bellwied, Fernando Flor, and Gabrielle Olinger. This material is based upon work supported by the
National Science Foundation under grant no. PHY1654219 and by the U.S. Department
of Energy, Office of Science, Office of Nuclear Physics,
within the framework of the Beam Energy Scan Topical (BEST) Collaboration. 
We also acknowledge the
support from the Center of Advanced Computing and
Data Systems at the University of Houston. J.N.H. acknowledges support from the Alfred
P Sloan Fellowship and the US-DOE Nuclear Science
Grant No. de-sc0019175. P.P. also acknowledges support by the DFG grant SFB/TR55.

\appendix

\section{Graphs}\label{app:graphs}
\begin{figure*}[h!]
\includegraphics[width=\textwidth]{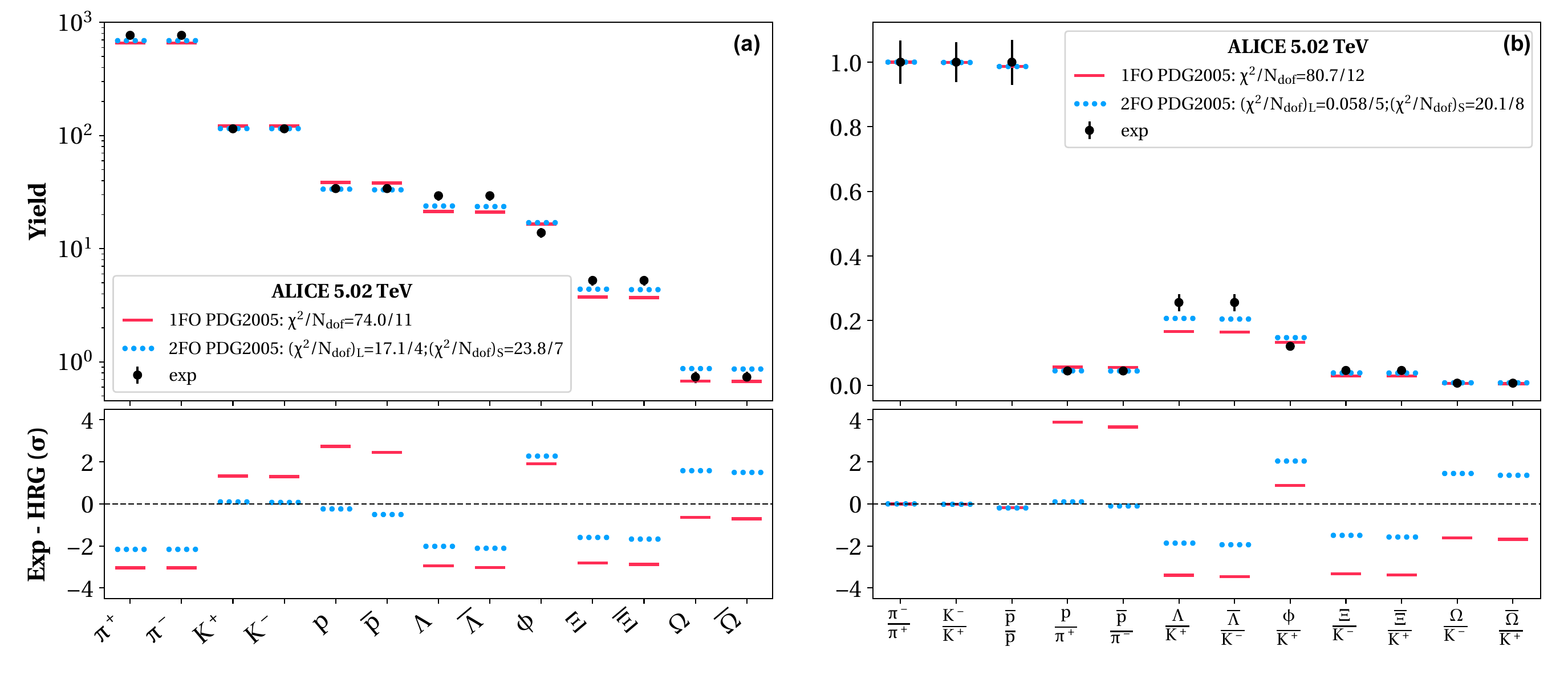}  \includegraphics[width=\textwidth]{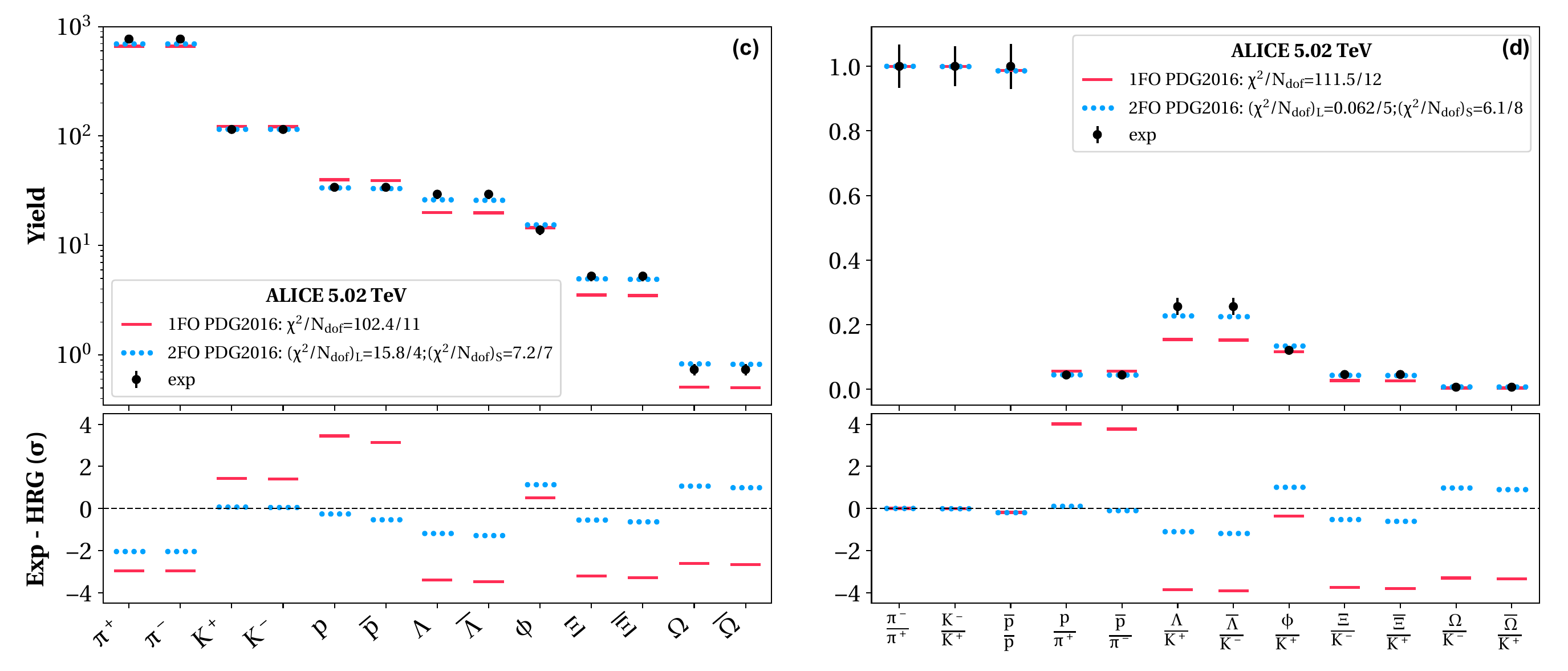} 
\caption{
Color online: ALICE PbPb $\sqrt{s_{NN}}=5.02$ TeV  data for particle yields (left) and 
ratios (right) in $0-10\%$ collisions, in comparison to HRG model calculations with the 
PDG2005 (upper) and PDG2016 (lower); deviations in units of experimental errors 
$\sigma$ are shown below each panel.
} 
\label{fit-yields_lhc1}
\end{figure*}

\begin{figure*}
\includegraphics[width=\textwidth]{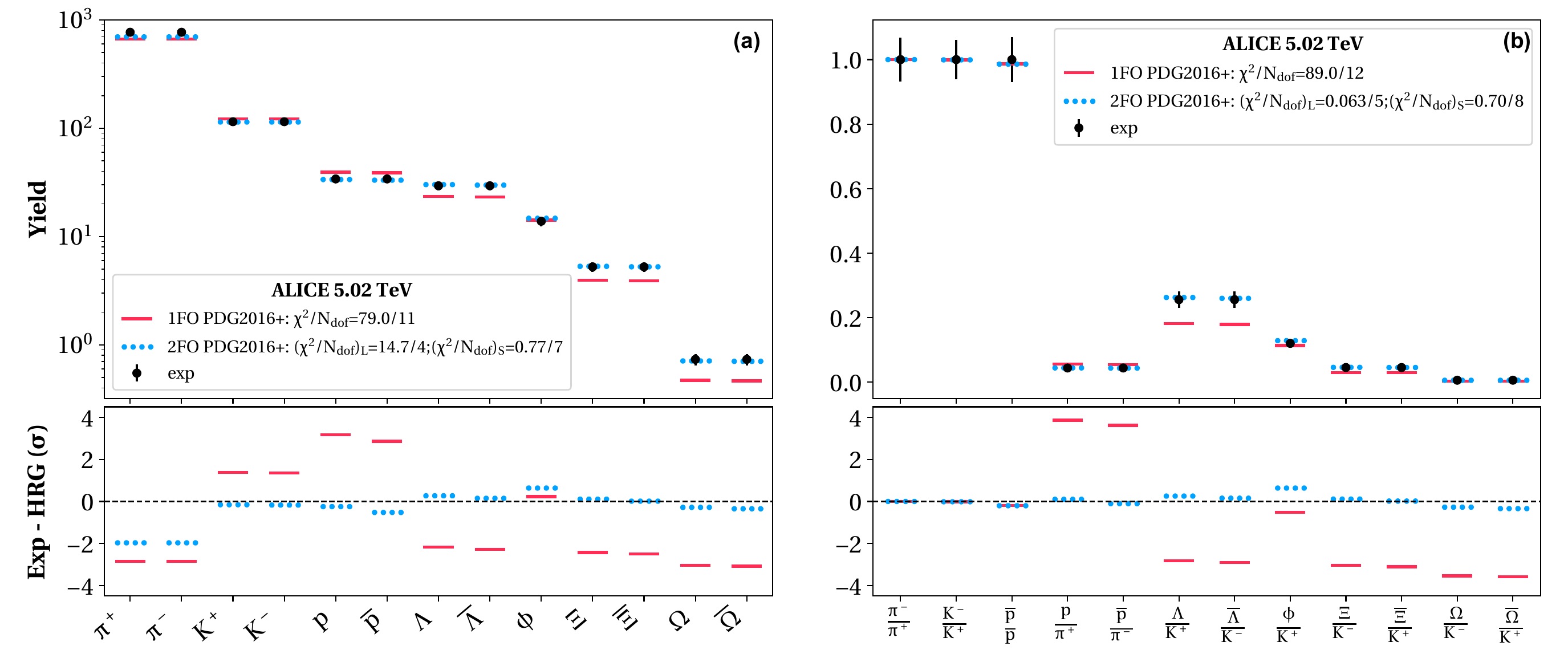}  \includegraphics[width=\textwidth]{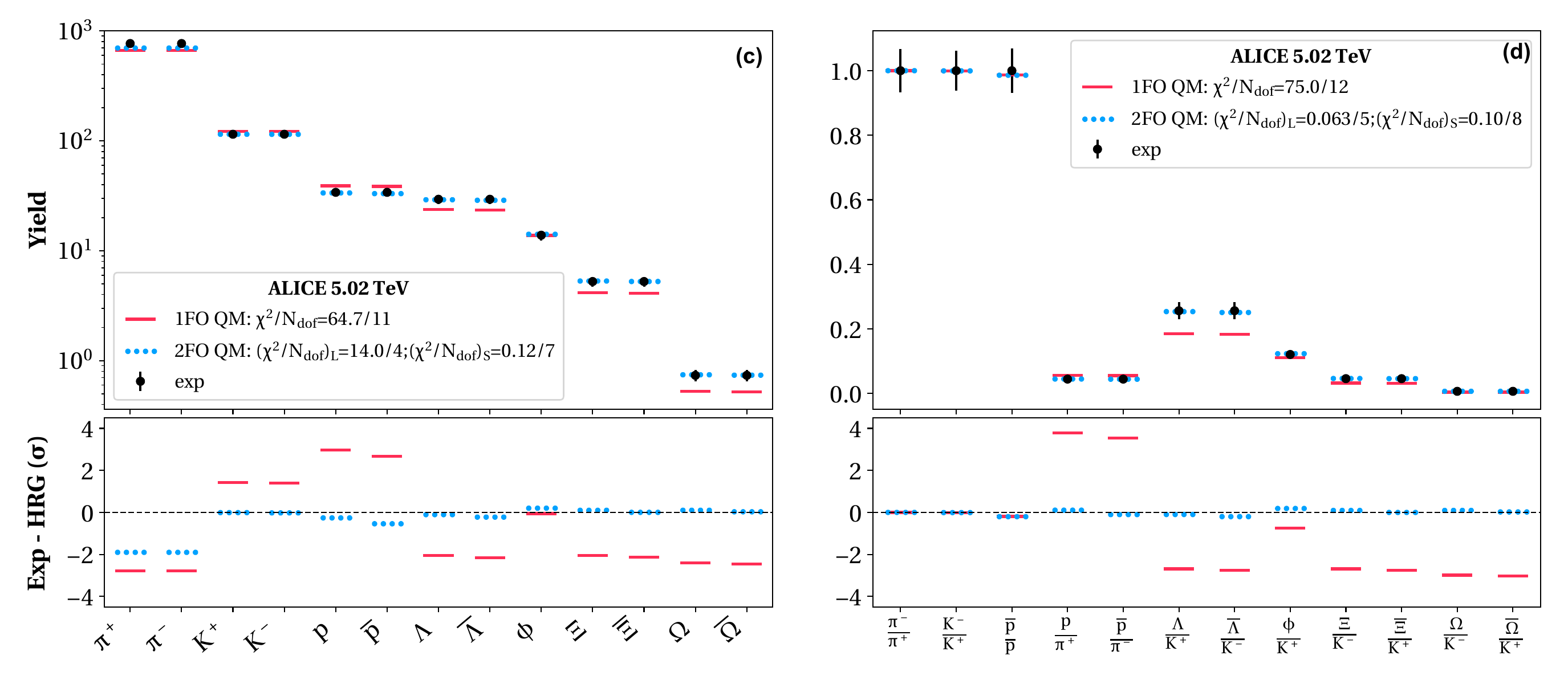}
\caption{
Color online: ALICE PbPb $\sqrt{s_{NN}}=5.02$ TeV  data for particle yields (left) and 
ratios (right) in $0-10\%$ collisions, in comparison to HRG model calculations with the 
PDG2016+ (upper) and QM (lower) lists; deviations in units of experimental errors 
$\sigma$ are shown below each panel.
} 
\label{fit-yields_lhc2}
\end{figure*}

\begin{figure*}[t]
\includegraphics[width=\textwidth]{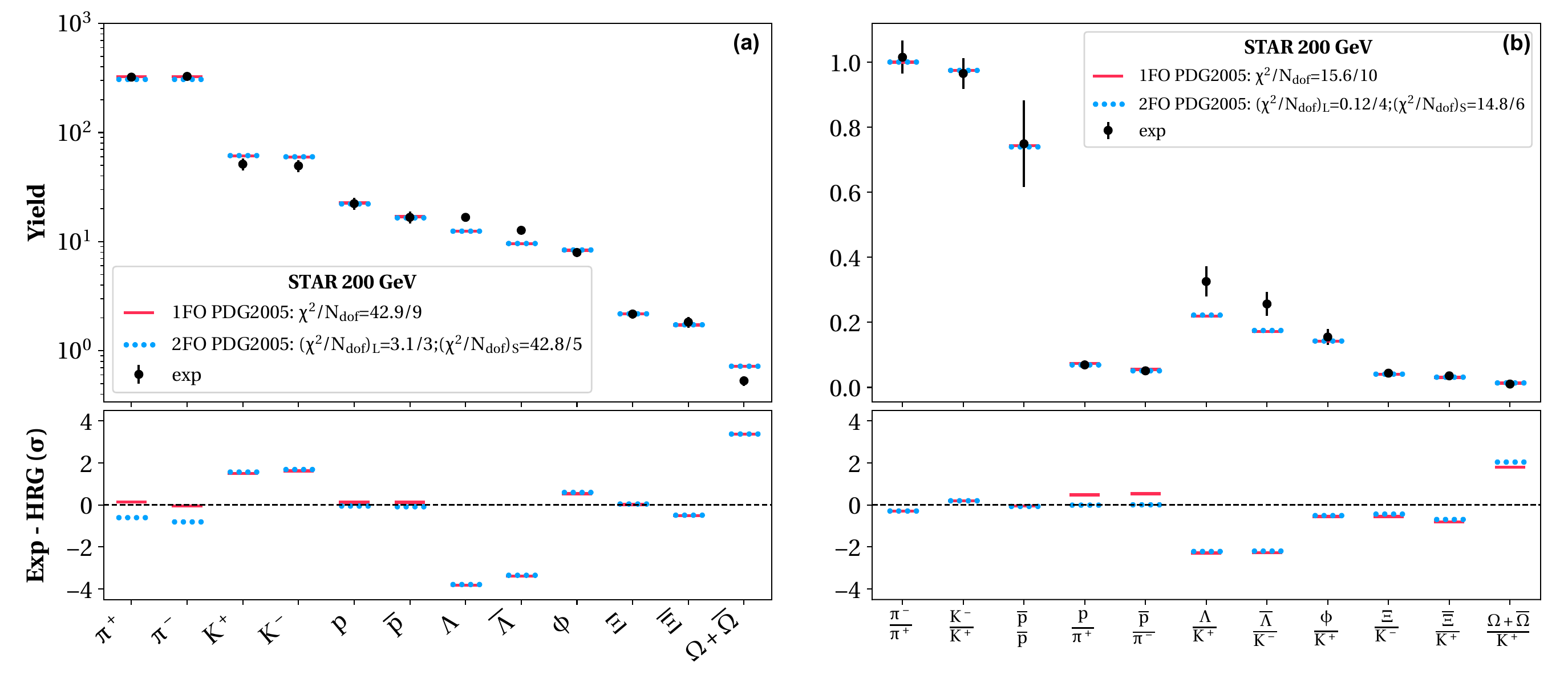}  
\includegraphics[width=\textwidth]{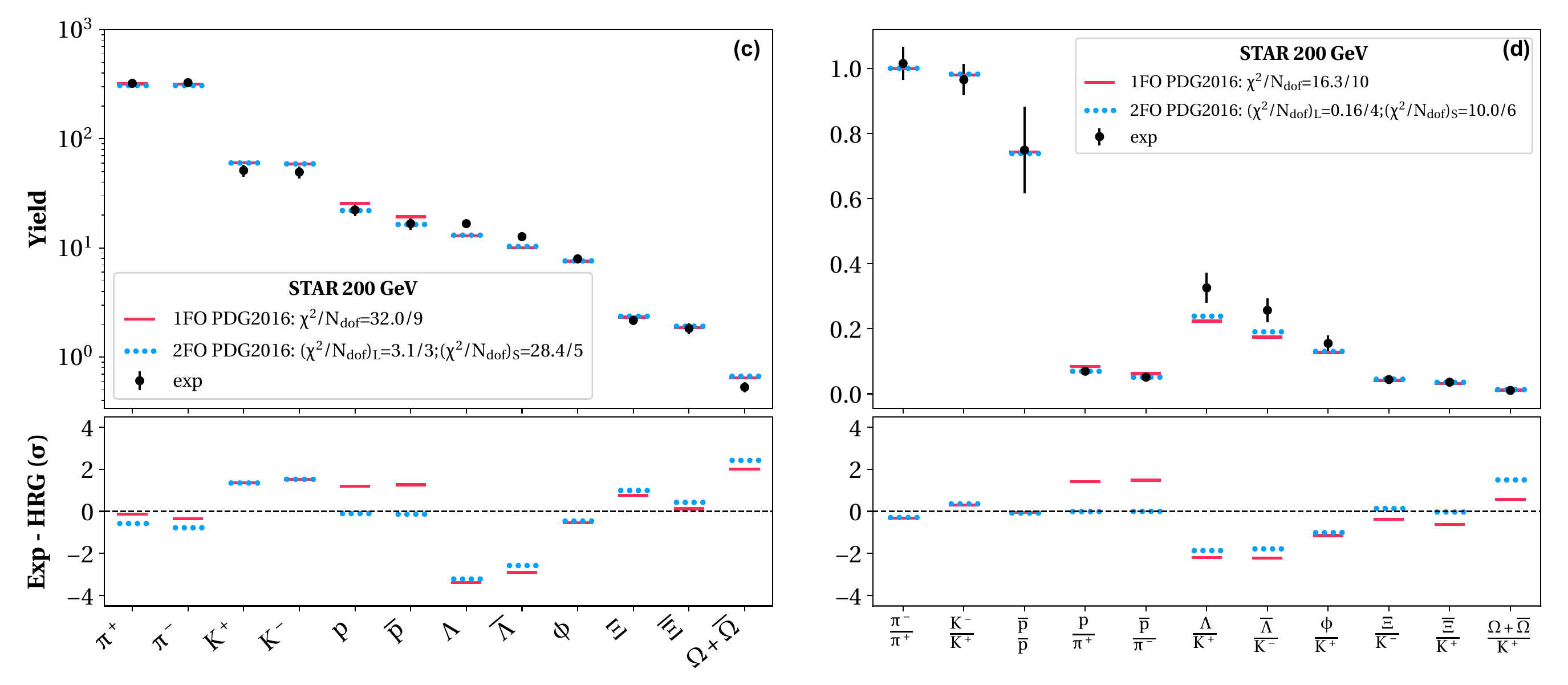}
\caption{
Color online: STAR AuAu $\sqrt{s_{NN}}=200$ GeV data for particle yields (left) and ratios 
(right) in $0-5\%$ collisions, in comparison to HRG model calculations with the PDG2005 
(upper) and PDG2016 (lower); deviations in units of experimental errors $\sigma$ are 
shown below each panel.
} 
\label{fit-yields_star1}
\end{figure*}

\begin{figure*}[t]
\includegraphics[width=\textwidth]{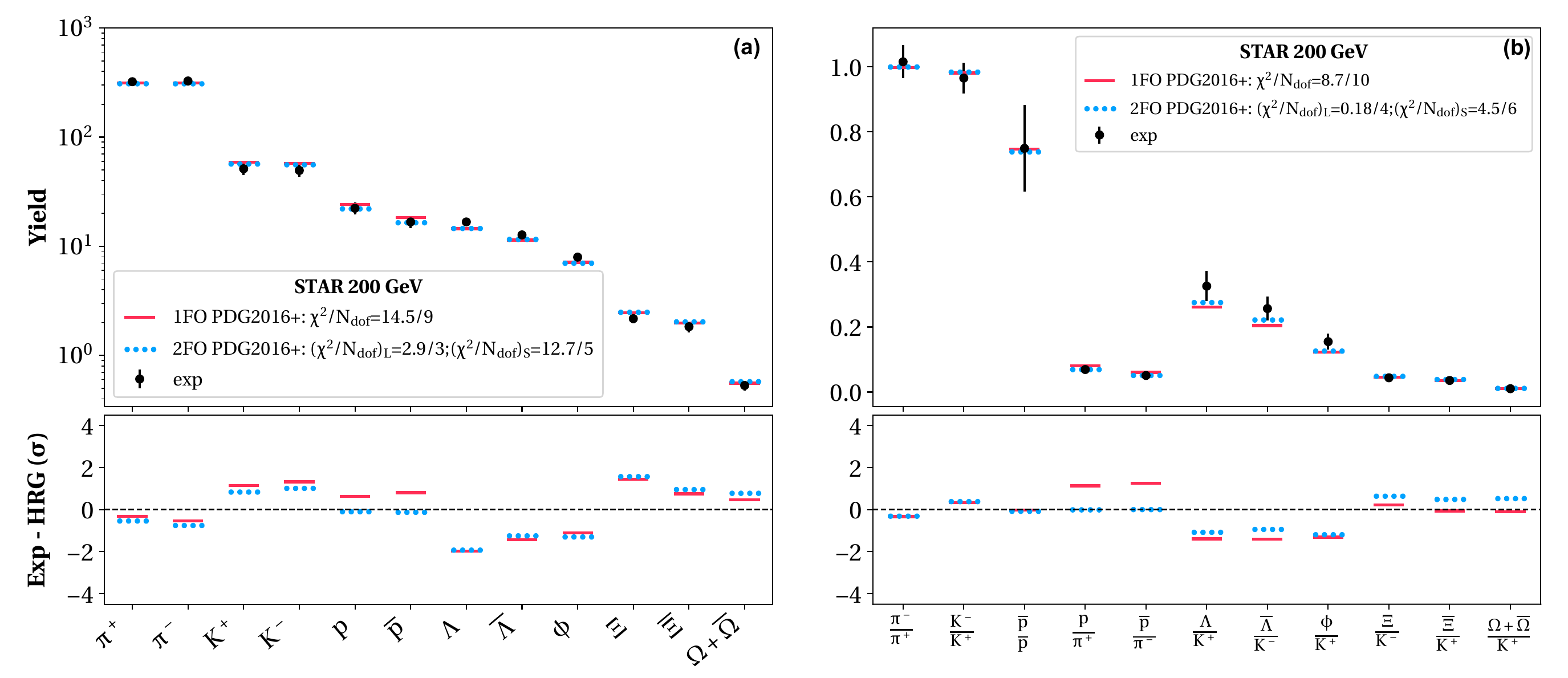}  \includegraphics[width=\textwidth]{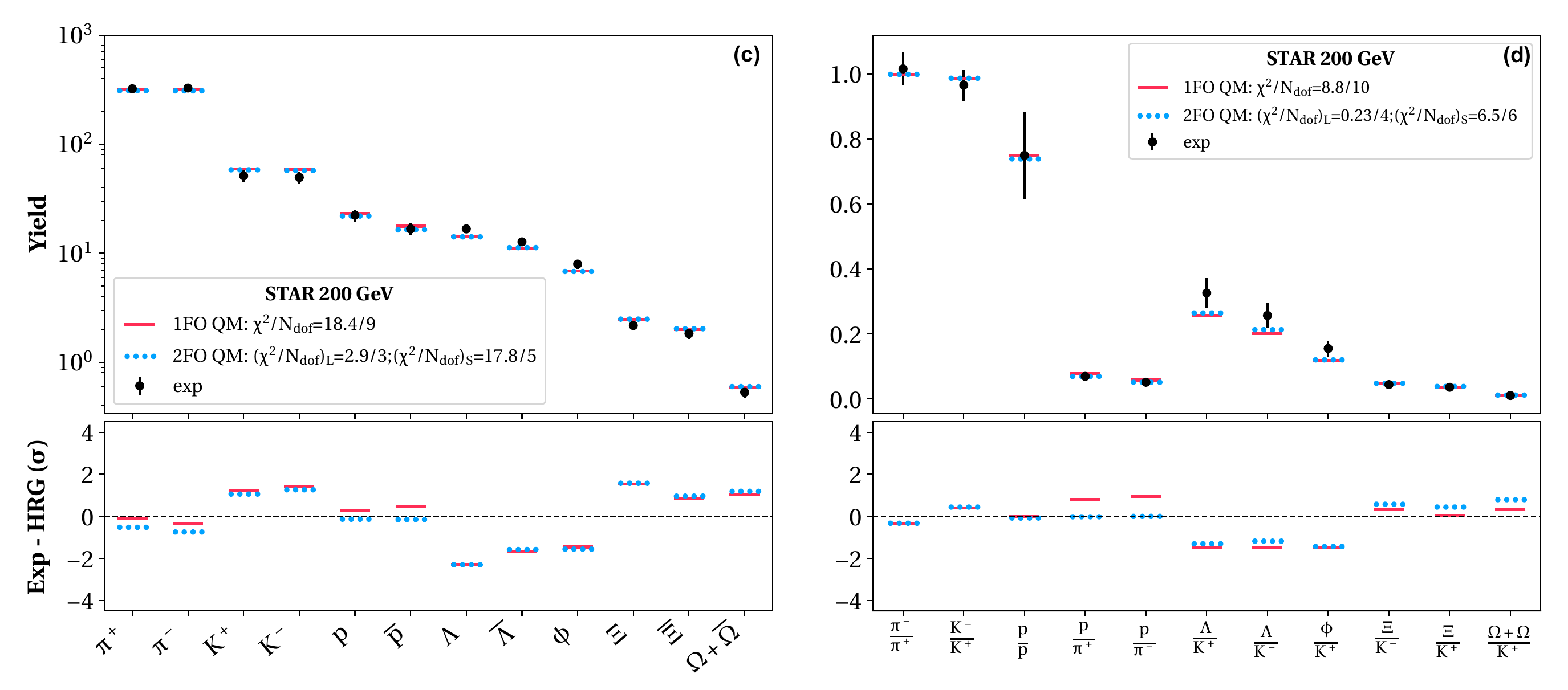}
\caption{
Color online: STAR AuAu $\sqrt{s_{NN}}=200$ GeV data for particle yields (left) and ratios 
(right) in $0-5\%$ collisions, in comparison to HRG model calculations with the PDG2016+ 
(upper) and QM (lower); deviations in units of experimental errors $\sigma$ are 
shown below each panel.
} 
\label{fit-yields_star2}
\end{figure*}

In this Appendix we show a comparison 
of the yields and ratios to experimental data.  The four figures (Figs.~\ref{fit-yields_lhc1}-\ref{fit-yields_star2}) show the four different particle list fits to yields and ratios at the LHC and RHIC, both in the single (red lines) and double (blue dotted lines) freeze-out scenarios.
In the former, $\Lambda$'s and $\Omega$'s tend to be under-predicted by 
thermal fits while protons tend to be over-predicted. The use of two separate freeze-out 
temperatures helps to alleviate this tension and tends to achieve a better overall fit of the data. The inclusion of more resonances does not solve the tension in the single freeze-out scenario,
as discussed above.


\bibliography{all}

\end{document}